\documentclass[prl,twocolumn,showpacs,
superscriptaddress,floatfix]{revtex4}
\usepackage{graphicx}

\begin{document}

\title{Mesoscopic Correlation with 
Polarization of Electromagnetic Waves}

\author{A.A.~Chabanov} 
\altaffiliation{Present address: CEMS, 
University of Minnesota, Minneapolis, MN 55545, USA}
\affiliation{Department of Physics, 
Queens College of the City University of New York, 
Flushing, NY 11367, USA}

\author{N.P.~Tr\'egour\`es} 
\affiliation{Laboratoire de Physique et Mod\'elisation 
des Milieux Condens\'es, 
CNRS/Universit\'{e} Joseph Fourier, Maison des Magist\`eres, 
B.P. 166, 38042 Grenoble, France}

\author{B.A.~van~Tiggelen} 
\affiliation{Laboratoire de Physique et Mod\'elisation 
des Milieux Condens\'es, 
CNRS/Universit\'{e} Joseph Fourier, Maison des Magist\`eres, 
B.P. 166, 38042 Grenoble, France}

\author{A.Z.~Genack} 
\affiliation{Department of Physics, 
Queens College of the City University of New York, 
Flushing, NY 11367, USA}

\date{September 4, 2003}

\begin{abstract}
Mesoscopic correlations are observed in the polarization 
of microwave radiation transmitted through a random waveguide. 
These measurements, supported by diagrammatic theory, permit 
the unambiguous identification of short, long, and infinite 
range components in the intensity correlation function, as 
well as an additional frequency-independent component.
\end{abstract}


\maketitle 
Mesoscopic transport of both classical and 
quantum mechanical waves is characterized by the degree 
of nonlocal intensity or current correlation which 
reflects the closeness to the Anderson localization 
threshold \cite{ShengBook,Mesobook}. Correlation is 
generally treated using scalar waves with the 
electromagnetic (EM) polarization or electron 
spin accounted for by doubling the density of 
states. The vector nature of EM radiation is 
exhibited in coherent backscattering \cite{CBS}, 
the photon Hall effect \cite{PHE}, and the memory 
of rotation of incident polarization \cite{Freund90}. 
These polarization effects reflect short-range field 
correlation and can be obtained within the field 
factorization approximation applicable to Gaussian 
statistics, as opposed to mesoscopic intensity 
correlation.

In this Letter, we present measurements and calculations 
of intensity correlation with polarization rotation that 
go beyond the field factorization approximation. 
Polarization is thereby shown to be an essential 
mesoscopic variable complementary to space, 
frequency, and time. Unlike these variables, 
however, the polarization variable is of finite 
range with only two independent values. Consequently, 
the field correlation function with polarization 
rotation has a simple form independent of scattering 
strength. It is determined exclusively by the cosine 
of the angular shifts in the polarization of the 
electric field at the source and detector and vanishes 
when either polarization is shifted by 90$^{0}$. 
This enables an unambiguous separation of the 
intensity correlation function, $C$, into short 
($C_1$) \cite{Shapiro86,Freund88}, long ($C_2$) 
\cite{Stephen87,Genack90}, and infinite ($C_3$) 
\cite{UCF,Feng88,C3} range components, which 
determine fluctuations in intensity, total 
transmission, and conductance, respectively. 
The frequency correlation function of each of 
these universal components can thereby be found. 
In addition, the presence of a frequency-independent 
component is identified. This term may correspond to 
a nonuniversal ``$C_{0}$" contribution, which reflects 
the scattering environment at the input and output 
surfaces \cite{C0}.

We begin with the conjecture, borne out by 
diagrammatic calculations and measurements 
reported below, that the structure of the 
correlation function of normalized intensity 
with regard to rotations in the polarization 
of source and detector, $C(\Delta\theta_{S}, 
\Delta\theta_{D})$, is analogous to that of 
the correlation function with displacement 
of the source and detector, $C(\Delta{\bf r}_{S},
\Delta{\bf r}_{D})$ \cite{Pnini}. In each component, 
$C$ depends upon a given degree of freedom only 
through its dependence upon the square of the field 
correlation function, $F\equiv |F_{E}|^{2}$. 
When expressed in terms of polarization 
rotations of the source and detector, this gives,
\begin{eqnarray}
C(\Delta\theta_{S},\Delta\theta_{D})&=&
F(\Delta\theta_{S})F(\Delta\theta_{D}) \nonumber \\ 
&+& A^{\prime}_{2}[F(\Delta\theta_{S}) 
+ F(\Delta\theta_{D})]  \nonumber \\ 
&+& A^{\prime}_{3}[1 + F(\Delta\theta_{S}) 
+ F(\Delta\theta_{D})   \nonumber \\ 
&\ \ \ &\ \  \ \  + \ 
F(\Delta\theta_{S})F(\Delta\theta_{D})]\, .
\label{cpol}
\end{eqnarray}
The leading contributions to $A^{\prime}_{2}$ and 
$A^{\prime}_{3}$ are of order $1/g$ and $1/g^{2}$, 
respectively, where $g$ is the dimensionless 
conductance.

The field correlation function with polarization 
rotation can readily be found for multiply scattered 
waves when the transmitted field is completely 
depolarized. In this case, the average intensity 
is independent of the polarization of the field 
at the source or detector and cross-polarized 
fields are uncorrelated. When the polarization 
of the detector is rotated by $\Delta\theta_{D}$ 
from an initial direction $\theta_{D}$, the detected 
field ${\bf E}(\theta_{D} + \Delta\theta_{D})$ may 
be expressed in terms of the vector sum of the field 
along $\theta_{D}$, ${\bf E}(\theta_{D})$, and the 
field perpendicular to this direction, 
${\bf E}(\theta_{D}+90^{0})$, as follows, 
${\bf E}(\theta_{D}+\Delta\theta_{D}) = 
{\bf E}(\theta_{D})\cos\Delta\theta_{D} + 
{\bf E}(\theta_{D}+90^{0})\sin\Delta\theta_{D}$. 
The field correlation function for the normalized 
average intensity, 
$\langle|{\bf E}(\theta_{D})|^{2}\rangle = 1$, is 
therefore, $F_{E}(\Delta\theta_{D}) \equiv 
\langle{\bf E}(\theta_{D}){\bf E}^{*}(\theta_{D}+
\Delta\theta_{D})\rangle = \cos\Delta\theta_{D}$. 
Here $\langle ... \rangle$ denotes the average 
over an ensemble of realizations. A similar 
argument gives the field correlation function of 
the detected field when the source polarization is 
rotated by $\Delta\theta_{S}$, $F_{E}(\Delta\theta_{S}) 
= \cos\Delta\theta_{S}$. The total field correlation 
function with polarization rotation is therefore given 
by $F_{E}(\Delta\theta_{S}, \Delta\theta_{D}) = 
F_{E}(\Delta\theta_{S}) F_{E}(\Delta\theta_{D}) = 
\cos\Delta\theta_{S} \cos\Delta\theta_{D}$.

The contributions to $C$ in Eq.~(1) can be expressed in 
terms of the product or sum of $F(\Delta\theta_{S})$ 
and $F(\Delta\theta_{D})$ or of a constant. Since these 
three types of terms are associated, respectively, with 
short, long, and infinite range correlation, it is convenient 
to group these terms separately. We therefore write,
\begin{eqnarray}
C(\Delta\theta_{S}, \Delta\theta_{D}) &=& (1 + A_{1})
\cos^{2}\!\Delta\theta_{S} \cos^{2}\!\Delta\theta_{D} 
\nonumber \\ &+& A_{2}\!\left(\cos^{2}\!\Delta\theta_{S} 
+ \cos^{2}\!\Delta\theta_{D}\right)+A_{3}\, .
\label{cpol}
\end{eqnarray}
For the case, in which nonuniversal ``$C_0$" contributions 
to $C$ are negligible, $A_{1}=A^{\prime}_{3}$, 
$A_{2}=A^{\prime}_{2}+ A^{\prime}_{3} $, and 
$A_{3}=A^{\prime}_{3}$. The analysis of Ref.~\cite{Pnini} 
has been generalized to vector waves in a detailed 
diagrammatic study \cite{PhDnicolas}. The relation 
$A_{1}=A_{3}$, which follows from Eq.~(1), can be 
understood from the topological structure of the 
underlying diagrams, and the equal weights of the 
two $A_{2}$ terms follow from reciprocity. In the 
diffusive regime in the absence of absorption, 
to order $1/g^2$,
\begin{equation}
A^{\prime}_{2}=\frac 23 {1\over g}\, ,\ \ \ 
A^{\prime}_3=\frac 2{15} {1\over g^2}\, ,
\label{1d}
\end{equation}
where $g=2\times Ak^2\ell/3\pi L$, which includes both 
polarizations. Here, $k$ is the wave number, $\ell$ 
is the transport mean free path, $A$ and $L$ are the 
cross-section area and length of the sample, respectively. 
Values of $A^{\prime}_{2}$ and $A^{\prime}_{3}$ slightly 
depend on absorption \cite{Pnini,Brouwer98,Rossum}. 
In our experiment, $L/L_{a}=3.6$, where $L_{a}$ is the 
absorption length. This gives $A^{\prime}_{2}=0.57/g$ 
and $A^{\prime}_{3} = 0.113/g^{2}$.

Measurements of the dependence of the field and intensity 
correlation functions upon polarization rotation for 
microwave radiation transmitted through a random 
dielectric sample are made with identical conical horns 
positioned 40 cm in front and behind the sample 
(Fig.~1). 
\begin{figure}
\includegraphics [width=\columnwidth] {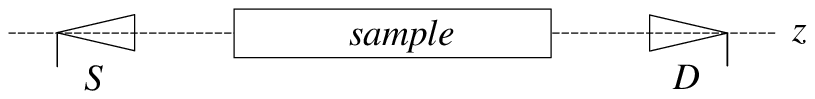}
\caption{Schematics of the experimental setup: 
\textit{S} (source) and \textit{D} (detector) 
microwave horns are positioned symmetrically 
in front and behind the sample and can be rotated
about their axes ($z$-axis) thus rotating the 
polarizations of the incident and detected waves.}
\end{figure}
Linearly polarized microwave radiation 
is launched from horn \textit{S} and a linearly 
polarized component of the transmitted field is 
captured by horn \textit{D}. The polarization 
of incident and detected waves can be rotated in 
the $xy$-plane by rotating the corresponding horn 
about its axis ($z$-axis). Because of the conical 
shape of the horns, the angular distribution of 
the incident and detected radiation do not depend 
strongly on the rotation of the horns. As a result, 
the variation in the detected 
field at a given frequency for a given sample 
realization is mainly due to the shift in the 
polarization of the field at the source and 
detector. The sample is composed of 0.95-cm-diameter 
alumina spheres with refractive index 3.14 embedded 
within Styrofoam spheres of diameter 1.9 cm and 
refractive index 1.04. The sample with an alumina 
volume fraction of 0.068 is contained in a 
7.3-cm-diameter copper tube of length $L=90$ cm. 
Measurements for random ensembles are obtained by 
momentarily rotating the sample tube about its 
axis to create new sample realizations, after which
spectra are taken for each orientation of the 
polarizers. Measurements are carried 
out in an ensemble of 12,000 sample realizations 
for 2 orientations of the source polarizer rotated 
by 90$^{0}$ and 7 orientations of the detector 
polarizer rotated in steps of 15$^{0}$, and also 
in an ensemble of 32,000 realizations for 2 
orientations of each polarizer rotated by 90$^{0}$. 
The transmitted field is measured in steps of 1 MHz 
from 16.95 to 17.05 GHz using a Hewlett-Packard 8772C 
vector network analyzer. This frequency range is 
centered at the peak of the fourth Mie resonance 
of the alumina spheres \cite{Chabanov01} and is 
much narrower than the width of the resonance so 
that propagation parameters within this range are 
nearly constant. At $\nu=17$ GHz, the number of 
transverse channels in the sample tube is 
$N=Ak^2/2\pi=84$. The transport mean free path 
estimated from Mie theory is $\ell =2.34$ cm, 
giving $g \approx 4N\ell/3L=2.91$. A fit of 
diffusion theory to the measured field 
correlation function with frequency shift gives 
$D=8.0$ cm$^2$/ns for the diffusion coefficient 
and $L_{a}=24.9$ cm for the absorption length.

The real and imaginary parts of 
$F_{E}(0,\Delta\theta_{D})$ obtained by averaging over 
frequency for an ensemble of 12,000 sample configurations 
are displayed by the crosses in Figs.~2a and 2b, 
respectively. The dashed curves represent the predictions 
--- $\cos \Delta \theta_{D}$ for the real part and 0 for 
the imaginary part. The orientation of the horn 
polarizations for the initial and final data points 
are indicated by the arrows. In addition, we find 
$F_{E}(90,\Delta\theta_{D})$ to vanish for all the values 
of $\Delta\theta_{D}$. Though the measurements are 
consistent with the predicted behavior of $F_{E}$, there 
is an increasing systematic deviation from theory with 
increasing $\Delta\theta_{S}$ or $\Delta \theta_{D}$. 
This deviation arises because, in addition to rotation 
of the polarization, there is a residual variation of 
the spatial intensity distribution on the sample, as the 
horn is rotated.
\begin{figure}
\includegraphics [width=\columnwidth] {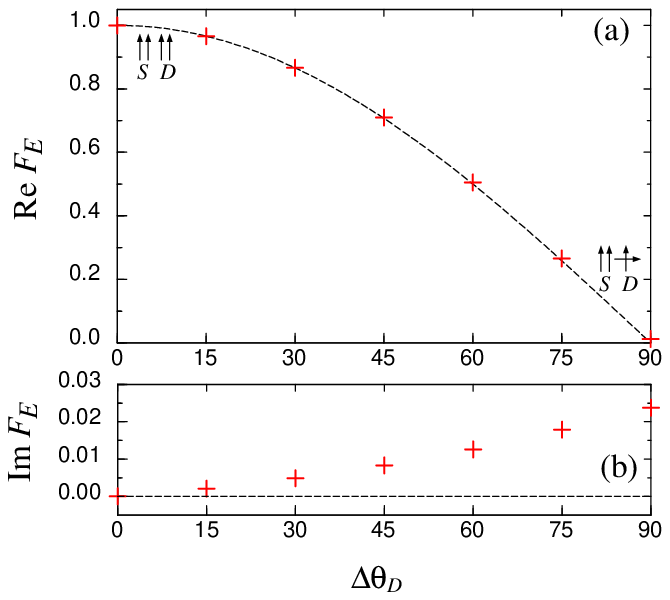}
\caption{(a) Real and (b) imaginary parts of the field 
correlation function $F_{E}(0,\Delta\theta_{D})$ plotted 
versus the angular shift of the polarization of the detected 
field, $\Delta\theta_{D}$, as the polarization of the incident 
wave is kept unchanged, $\Delta\theta_{S}=0$. The arrows 
indicate orientations of the source (\textit{S}) and detector 
(\textit{D}) polarizers for the initial and final data points. 
The dashed lines represent the theoretical prediction, 
$F_{E}(0,\Delta\theta_{D})=\cos(\Delta\theta_{D})$.}
\end{figure}

The intensity correlation functions $C(0,\Delta\theta_{D})$ 
and $C(90,\Delta\theta_{D})$ for the same data set are shown 
by the blue crosses in Figs.~3a and 3b, respectively. In 
addition, $F(\Delta\theta_{D})$ is shown in Fig.~3a by the 
red crosses, which lie close to the solid curve of 
$\cos^{2}\!\Delta\theta_{D}$. A Legendre polynomial fit of 
$C(0,\Delta\theta_{D})$ and $C(90,\Delta\theta_{D})$ shows that, 
with high accuracy, they are of the form $a+b\cos^{2}\!\Delta 
\theta_{D}$, where $a$ and $b$ are constants, in agreement 
with Eq.~(2). The dashed curve in Fig.~3a is a fit to the 
first 4 data points of $C(0,\Delta\theta_{D})$ of the function 
$a+b\cos^{2}\!\Delta\theta_{D}$, where $a=A_{2}+ A_{3}$ and 
$b=1+ A_{1}+ A_{2}$. Only the first four points are used because 
the systematic deviation in $C$ due to the redistribution 
of intensity with rotation cannot be ignored for larger 
polarization rotations. Though uncertainty is introduced into 
$C_1$ and $C_2$ by the horn rotation, there is no such effect 
on $C_3$ because it is independent of both polarization and 
displacement. Thus the measured value for $A_3$ is not affected 
by the horn rotation. It is found directly from $C(90,90)$ for 
an ensemble of 44,000 realizations, $A_{3}=0.029\pm 0.001$. 
From the fit, we obtain $A_{1}+ A_{2}=0.286\pm 0.005$ and 
$A_{2}+ A_{3}= 0.293\pm 0.004$. Hence $A_{2}=0.264 \pm 0.006$ 
and $A_{1}= 0.022 \pm 0.010$. The dashed curve in Fig.~3b is 
$A_{3}+A_{2}\cos^{2}\!\Delta \theta_{D}$ with the $A_{2}$ and 
$A_{3}$ found above. The difference between the experimental 
data points and the curve represents the loss of correlation 
due to the redistribution of intensity with the horn rotation.
\begin{figure}
\includegraphics [width=\columnwidth] {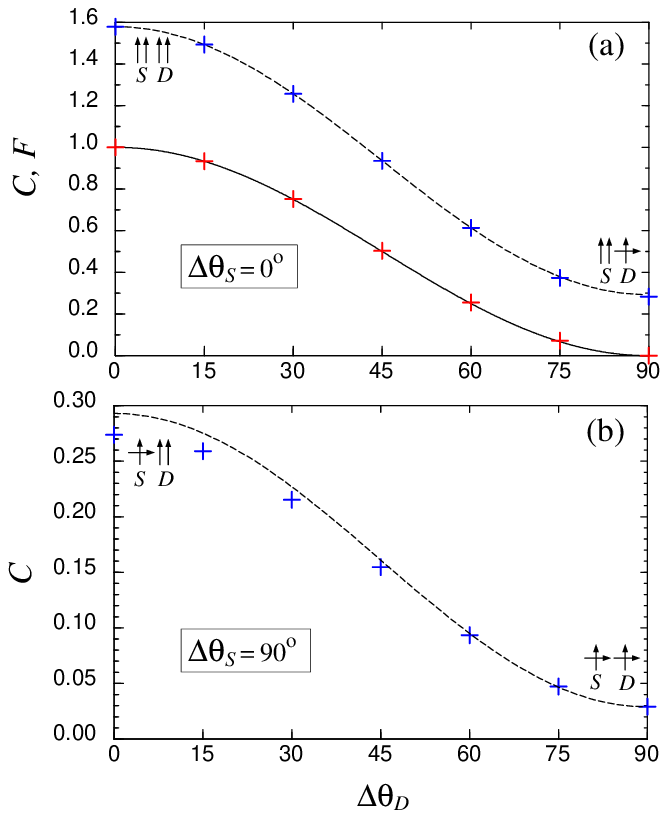}
\caption{(a) Intensity correlation functions $C$ and $F$ 
plotted versus $\Delta\theta_{D}$ for $\Delta\theta_{S}=0$. 
The solid line is the prediction, $F(\Delta\theta_{D})=
\cos^{2}(\Delta\theta_{D})$. The dashed line is the fit of 
$C(0,\Delta\theta_{D})$ to the first 4 data points. (b) $C$ 
versus $\Delta\theta_{D}$ for $\Delta\theta_{S}=90^{0}$. 
The dashed line is the prediction $A_{3}+A_{2}\cos^{2}\!\Delta 
\theta_{D}$ with the $A_{2}$ and $A_{3}$ found from the fit 
in (a).}
\end{figure}

The coefficients $A_{i}$ for arbitrary frequency shift can be 
determined from spectral measurements for three orientations 
of the source and detector as follows, $A_{3}=C(90,90)$, 
$A_{2}=C(0,90)-A_{3}$, and $A_{1}=C(0,0)-|F_E(0,0)|^2-2A_{2}-A_{3}$. 
We have utilized the above procedure in the ensemble of 44,000 
sample realizations, to find $A_i(\Delta\nu)$ for frequency shifts 
$\Delta\nu$ up to 100 MHz, that is almost 100 times the Thouless 
frequency $\Delta\nu_{Th}\equiv 6D/2\pi L^{2}\approx 1$ MHz. 
To compensate for the loss of correlation in $C(0,90)$ due to 
the redistribution of intensity, the corresponding frequency 
correlation function was multiplied by a constant to make it equal 
to the $A_{2}+ A_{3}$ at zero frequency shift. 
The resulting dependencies are shown by the squares in Fig.~4. 
$A_2(\Delta\nu)$ and $A_3(\Delta\nu)$ are predicted \cite{Rossum} 
to fall asymptotically as $A_2(\Delta\nu)\sim 1/(\Delta\nu)^{1/2}$ 
and $A_3(\Delta\nu)\sim 1/(\Delta\nu)^{3/2}$. We find, however, that 
$A_3$ exhibits a \emph{nonzero} correlation for $\Delta\nu\gg 
\Delta\nu_{Th}$, that largely exceeds error bars, and that 
$A_{2}$ is shifted from the predicted curve by a nearly constant value. 
Figs.~4a and 4b show that agreement with theory is obtained for 
$A_{2,3}$, if \emph{frequency-independent} contributions with the 
values $A_2(\infty)=-0.015$ and $A_3(\infty)=0.007$ are incorporated, 
respectively, in $A_2$ and $A_3$. We do not observe an asymptotic 
background in $A_1$ (Fig.~4c). The observed $A_1(\Delta\nu)$, however, 
falls faster than predicted by diagrammatic theory. To calculate 
$A_1(\Delta f)$, we have modified the theory for $A_3(\Delta f)$ 
\cite{Rossum}. 
\begin{figure}
\includegraphics [width=\columnwidth] {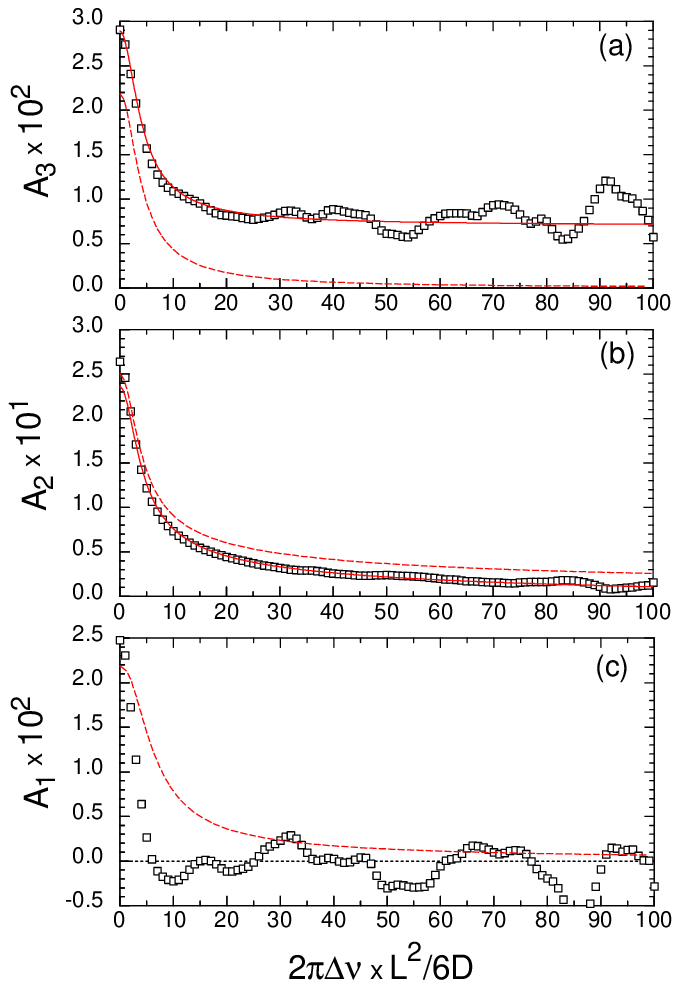}
\caption{Coefficients $A_{3}$ (a), $A_{2}$ (b), and $A_{1}$ (c) 
plotted versus frequency shift. The long-dashed curves represent 
predictions of mesoscopic theory [17] for $A_3$ (a), $A_2^\prime$ 
(b), and $A_1$ (c) with $g=2.27$ and $L/L_{a}=3.6$. The 
frequency-independent contributions with the values 
$A_3(\infty)=0.007$ and $A_2(\infty)=-0.015$ have been added, 
respectively, to the predicted $A_3$ and $A_2^\prime$ to obtain 
the solid curves and agreement with the data.}
\end{figure}

After subtracting the asymptotic values we find 
$A_3(0)-A_3(\infty)=A_3^\prime(0)=0.022\pm 0.001$ and 
$A_2(0)-A_2(\infty)=A_2^\prime(0)+A_3^\prime(0)=0.279\pm 0.004$ 
at zero frequency shift. Note that $A_3^\prime(0)=A_1(0)$ holds 
within the error bars. Using Eq.~(3) the value for $A_3^\prime(0)$ 
yields $g=2.27\pm 0.05$ for the dimensionless conductance. This 
differs from the estimate, reflecting effects of localization, 
internal reflection, and finite scatterer density. Using this value, 
Eq.~(3) predicts $A^{\prime}_{2}(0)=0.252$, and hence 
$A^{\prime}_{2}(0)+A^{\prime}_{3}(0)= 0.274$, which is also 
consistent with observations.

It is tempting to associate the frequency-independent background 
term with the $C_0$ term of Ref.~\cite{C0}. A preliminary calculation
for point dipoles in a tube geometry with $N$ transverse channels
\cite{Sergey} shows $C_0$ to be independent of frequency shift,
positive and of order $1/N = 0.012$, and with contributions to 
$A_2$ and $A_3$ and not to $A_1$. This is consistent with the 
observed value $A_3(\infty)=0.007$, but not with the \emph{negative} 
value $A_2(\infty)=-0.015$. A vector Mie theory for $C_0$ along with 
measurements in samples with different longitudinal and transverse 
dimensions may shed light on this issue.

In conclusion, mesoscopic correlations have been found in the 
polarization of microwave radiation transmitted through 
a random waveguide. The intensity correlation 
function with polarization rotation is shown to be 
composed of the product and sum of the square of the 
field correlation function and of a constant term. 
These three types of terms are associated, respectively, 
with short, long, and infinite range correlation. The variation 
with frequency shift of each of these terms has been measured. 
The long and infinite range terms are found to include a 
frequency-independent component, which may be related to 
the short-duration interaction of the wave with regions 
of limited spatial extent near the input and output surfaces 
of the sample. The simple form of the three contributions to 
the intensity correlation function with polarization rotation 
allows an unambiguous separation of correlation into short, 
long, and infinite range contributions. This method is readily 
applicable to optical measurements and can be used to determine 
the degree of intensity correlation in a sample and hence the 
closeness to the localization threshold. Analogous correlations 
can also be expected in electron spin propagating in a random 
potential. 

It is a pleasure to acknowledge stimulating discussions with
R.~Pnini, S.E.~Skipetrov, and B.~Shapiro. We would also like to
thank G.~German and Z.~Ozimkowski for help in constructing the experimental 
apparatus, as well as K.~Chabanov and B.~Hu for experimental assistance. 
This research is sponsored by the U.S. Army Research Office
(DAAD190010362), the National Science Foundation (DMR0205186), and
GDR 2253 IMCODE of the CNRS.

\end{document}